\documentclass[a4paper,11pt]{article}
\usepackage{pos}
\usepackage[utf8]{inputenc}
\usepackage{amsmath}
\usepackage{float}
\usepackage{tabulary}
\usepackage{empheq}
\usepackage{tabu}

\def\Q {$\mathcal{Q}_{\rm CDM}$~}
\def\non{\nonumber\\}
\newcommand{\code}[1]{\texttt{#1}}
\DeclareUnicodeCharacter{2212}{-}

\title{Cosmological tensions and \Q as an alternative to $\Lambda$CDM}
\author[a]{Amin Aboubrahim}
\author*[b]{Pran Nath}

\affiliation[a]{Department of Physics and Astronomy, Union College, \\
807 Union Street, Schenectady, NY 12308, U.S.A.}

\affiliation[b]{Department of Physics, Northeastern University, \\
111 Forsyth Street, Boston, MA 02115-5000, U.S.A.}

\emailAdd{abouibra@union.edu}
\emailAdd{p.nath@northeastern.edu}

\abstract{The Standard Model of cosmology, $\Lambda$CDM, while enormously successful, is currently
 unable to account for several cosmological anomalies the most prominent of which are in the measurements of the Hubble parameter and $S_8$. Additionally, the inclusion of the cosmological
 constant is theoretically unappealing. This has lead to extensions of the model such as the
 use of fluid equations for interacting dark matter and dark energy which, however, are ad hoc since they do not appear to arise from a Lagrangian. Recently, we have proposed 
 \Q as an alternative to $\Lambda$CDM which is a dynamical model of a quintessence field interacting with dark matter within a field theoretic approach. In this approach, we analyze the effect of the dark matter mass, the dark matter-dark energy interaction strength and the dark matter self-interaction on the cosmological parameters. Further, within \Q  we 
  investigate the possible alleviation of the Hubble tension and the $S_8$ anomaly and the nature of dark energy.}

\FullConference{Proceedings of the Corfu Summer Institute 2024 "School and Workshops on Elementary Particle Physics and Gravity" (CORFU2024)\
12 - 26 May, and 25 August - 27 September, 2024\\
Corfu, Greece\\}

\begin{document}
\maketitle

\section{Introduction} \label{sec:int}

The $\Lambda$CDM model has been successful  in explaining a wide range of cosmological data  but has recently faced challenges in explaining some anomalies, the most prominent of which are a discrepancy in measurements of the Hubble parameter (known as the Hubble tension)
and $S_8$. To take account of the latter, many works  directly use  continuity equations for dark matter and dark energy with sources chosen to satisfy the overall 
energy conservation. However, such approaches do not have any fundamental origin.
More specifically, the fluid equations currently in use have no simple field-theoretic basis.  In this 
note, we discuss a field-theoretic formulation of interacting dark matter and dark energy proposed
in ref.~\cite{Aboubrahim:2024spa} which is fully field-theoretic as an alternative to 
$\Lambda$CDM. For specificity we will discuss an explicit model of two interacting spin zero fields,
one for dark matter and the other for dark energy. In this framework we 
compute the background and linear perturbation equations 
within this field-theoretic formalism. We then carry out numerical fits to the cosmological data which include data from Planck~\cite{Aghanim:2018eyx} (with lensing), BAO, Pantheon, SH0ES, and WiggleZ, and specifically discuss the $H_0$ and $S_8$ tensions.  At the end we will summarize our results and draw some conclusions.

\section{$\Lambda$CDM and the fluid equations}

The Lagrangian of  $\Lambda$CDM is given by ${\cal L}= \frac{1}{16\pi G}(R-2 \Lambda)+ {\cal L}_{\rm CDM}$ and contains no direct interaction between dark matter and dark energy.
 In models of quintessence, the $\Lambda$ term is replaced by a dynamical spin zero field. 
 In the so-called two-fluid model, one adopts the continuity equations for dark matter and
 dark energy  with the inclusion of source terms which are chosen so that the total energy is conserved.  Often the fluid equations do not specifically refer to fields but deal directly with 
 energy densities and the equations of state. However, it is important to consider what 
 the assumptions of the fluid equations are by looking at their explicit form within a field-theoretic formulation.  Thus, we consider two spin zero fields, one of which is a quintessence
 dark energy field and the other a dark matter field and assume there is a coupling 
 between them. In this case the continuity equations that follow from them are given by
    \begin{align}
 {\cal D}_\alpha T_\phi^{\alpha \beta} &= J_\phi^\beta,~~\text{(DE)}; ~~
  {\cal D}_\alpha T_\chi^{\alpha \beta} = J_\chi^\beta, ~~\text{(DM)}.  
  \label{fluid}
 \end{align} 
 In a Lagrangian theory with an interaction between the $\phi$ and $\chi$ fields, 
 the source terms $J_\phi^\beta$ and $J_\chi^\beta$ are determined by the Lagrangian
 equations of motion and in general do not satisfy the simple relation $J_\phi^\beta =-  J_\chi^\beta$.
 However, the conservation of energy-momentum is automatic in a Lagrangian formulation
 and does not need to be imposed by hand.
  We contrast this with the fluid equations currently in use where, one assumes no underlying 
  Lagrangian, but adopts the expressions of Eq.~(\ref{fluid}) with the
   constraint $J_\phi^\beta =-  J_\chi^\beta$ which is introduced in an ad hoc manner. On the other hand all the fundamental theories of physics are based on Lagrangians and  an action principle. This includes the Standard Model of particle physics, Einstein theory, and string theory.  So the 
   fluid model as currently used cannot be considered a  fundamental  cosmological model.

\section{\Q as an alternative to $\Lambda$CDM}

As mentioned earlier we discuss now a Lagrangian formulation of interacting 
    dark matter (DM) and dark energy (DE) as an alternative to $\Lambda$CDM.  We will use a specific model of DM and DE as an illustrative example but the underlying formalism is valid for any field-theoretic choice of DM and DE. Thus as an  illustrative example of  \Q we consider a Lagrangian formulation of interacting two spin zero DM and DE  fields  whose action and total potential are given by      
\begin{align}
A&= \int \text{d}^4 x\,\sqrt{-g} \left[-\frac{1}{2}\phi^{,\mu} \phi_{,\mu} - \frac{1}{2} \chi^{,\mu} \chi_{,\mu} -V(\phi, \chi)\right], \\        
    V(\phi,\chi)&=  V_1(\chi)+ V_2(\phi) + V_3(\phi,\chi).          
\end{align} 
This is a fairly general form of the Lagrangian for two  interacting spin zero fields.  We will specify the forms of the potentials when we carry out the numerical analysis later. 
For the background we consider a flat, homogeneous and isotropic universe characterized by the Friedmann-Roberston-Walker (FRW) metric written in conformal time so that 
\begin{equation}
    \text{d}s^2=g_{\mu\nu}\text{d}x^\mu \text{d}x^\nu=a^2(-\text{d}\tau^2+\gamma_{ij}\text{d}x^i \text{d}x^j),
\end{equation}
 where $a$ is time-dependent scale factor;  $\gamma_{ij}$ are spatial components of the metric; and  $\tau$ is the conformal time, so that $\text{d}\tau=\text{d}t/a(t)$.
 The background fields satisfy the following KG equations 
 \begin{align}
&\chi_0^{\prime\prime}+2\mathcal{H}\chi_0^\prime+a^2(\Bar{V}_1+\Bar{V}_3)_{,\chi}=0, 
~\text{and}~~\phi_0^{\prime\prime}+2\mathcal{H}\phi_0^\prime+a^2(\Bar{V}_2+\Bar{V}_3)_{,\phi}=0,
\end{align}
where $\bar V(\phi, \chi)\equiv V(\phi_0, \chi_0)$ and $\bar V_{1,\chi} \equiv (V_{1,\chi})_{\chi=\chi_0}$, etc; and $\mathcal{H}= a H$, with $H=\dot a /a$.
The field theory model  gives the following DE and DM continuity equations
\begin{align}
&\rho^\prime_\phi+3\mathcal{H}(1+w_\phi)\rho_\phi=Q_\phi\,, ~~\text{(DE)}\\
&\rho^\prime_\chi+3\mathcal{H}(1+w_\chi)\rho_\chi=Q_\chi\,,~~\text{(DM)}
\end{align}
where the source terms are $Q_\phi=\Bar{V}_{3,\chi}\chi^\prime$ and $Q_\chi=\Bar{V}_{3,\phi}\phi^\prime$. Energy conservation requires that the total energy density is
\begin{align}
&\rho^\prime+3\mathcal{H}(1+w)\rho=0,
\end{align}
with $\rho$ defined in such a way to avoid double counting, i.e., $\rho=\rho_\phi+\rho_\chi-V_3$.

We note here that in the two-fluid model one sets $Q_\phi=-Q_\chi=Q$
  so that the fluid equations assume the form
\begin{align}
&\rho^\prime_\phi+3\mathcal{H}(1+w_\phi)\rho_\phi=Q\,, ~~~~\text{(DE)}\\
&\rho^\prime_\chi+3\mathcal{H}(1+w_\chi)\rho_\chi=-Q\,.~~\text{(DM)}.
\end{align}
While the assumption  $Q_\phi=-Q_\chi=Q$  guarantees energy conservation, the constraint $Q_\phi=-Q_\chi$  is ad hoc, and makes the model non-Lagrangian.

\section{Linear perturbations}

We discuss now linear perturbations around the background of the spin zero fields so that 
\begin{align}
\chi(t,\Vec{x})=\chi_0(t)+\chi_1(t,\Vec{x})+\cdots,~~\phi(t,\Vec{x})=\phi_0(t)+\phi_1(t,\Vec{x})+\cdots
\end{align}
Perturbations of the metric in a general gauge is given by
\begin{equation}
\label{gen}
    g^{00}=-a^{-2}(1-2A), 
   ~ g^{0i}=-a^{-2}B^i, 
    ~g^{ij}=a^{-2}(\gamma^{ij}-2H_L \gamma^{ij}-2H_T^{ij}),
\end{equation}
where $A$ is a scalar potential, $B^i$ a vector shift, $H_L$ is a scalar perturbation to the spatial curvature and  $H_T^{ij}$ is a trace-free distortion to the spatial metric. There are two types of gauges that are generally assumed in the analysis.
These are the synchronous and conformal (or Newtonian) gauges.  In the synchronous gauge $g^{00}$ and $g^{0i}$ are not perturbed and so the line element has the form: 
 $\text{d}s^2=a^2(\tau)\left[-\text{d}\tau^2+(\delta_{ij}+h_{ij})\text{d}x^i \text{d}x^j\right]$, while $A=B=0, ~H_L=\frac{1}{6}h$, where $h$ is trace of the metric perturbations $h_{ij}$.
The conformal (Newtonian) gauge is characterized by 
    $B=H_T=0,~A\equiv \Psi, ~H_L \equiv \Phi$.
Our analysis is carried out in a general gauge which can then reduce to either the
synchronous  gauge and or the conformal gauge based on the above prescription. 

The perturbations of the stress-energy tensor is given by
\begin{align}
T^{\mu}_{\nu}=\bar{T}^{\mu}_{\nu}+\delta T^{\mu}_{\nu},
\end{align}
 where the stress tensor in component form is given by 
\begin{align}
    T^{0}_0&=-\rho-\delta\rho, ~
    T^{0}_i=(\rho+p)(v_i-B_i),\\
    T^{i}_0&=-(\rho+p)v_i,
    ~T^{i}_j= (p+\delta p)\delta^i_j+p\Pi^i_j , 
\end{align}
with $\Pi^i_j$ representing the anisotropic stress, $v_i$ the 3-velocity,  $\delta\rho$ and $\delta p$ being the density and pressure perturbations. The off-diagonal element $ \delta T^0_i$ gives the velocity divergence so that 
\begin{align}
\delta T^0_i=-a^{-2}\phi_0^\prime \delta\phi_{,i},
\end{align}
where the velocity divergence $\theta$ is defined in Fourier space so that  
 $\theta=ik^i v_i$ or alternately $\Theta\equiv(1+w) \theta$ are determined by 
\begin{align}
    \rho_\phi \Theta_\phi=\frac{k}{a^2}\phi_0^\prime\phi_1,~\rho_\chi \Theta_\chi=\frac{k}{a^2}\chi_0^\prime\chi_1\,.  
\end{align}
The first order perturbations $\phi_1$ and $\chi_1$ are determined via the
solutions to the KG equations they satisfy which are 
\begin{align}
&\phi_1^{\prime\prime}+2\mathcal{H}\phi_1^\prime+(k^2+a^2\Bar{V}_{,\phi\phi})\phi_1+a^2\Bar{V}_{,\phi\chi}\chi_1+2a^2\Bar{V}_{,\phi}A+(3H_L^\prime-A^\prime+kB)\phi_0^\prime=0, \\
&\chi_1^{\prime\prime}+2\mathcal{H}\chi_1^\prime+(k^2+a^2\Bar{V}_{,\chi\chi})\chi_1+a^2\Bar{V}_{,\chi\phi}\phi_1+2a^2\Bar{V}_{,\chi}A+(3H_L^\prime-A^\prime+kB)\chi_0^\prime=0.
 \end{align}
Using the above, one computes the density perturbations defined so that
\begin{equation}
    \delta_i\equiv\frac{\delta\rho_i}{\bar{\rho}_i}=\frac{\rho_i(t,\Vec{x})-\bar{\rho}_i(t)}{\bar{\rho}_i},
\end{equation}
From here on we will assume specific forms of the potentials which are
\begin{align}
 ~V_1(\chi)&=\frac{1}{2} m_\chi^2 \chi^2+ \frac{\lambda}{4} \chi^4, ~~\text{(DM)},\\
~V_2(\phi)&= \mu^4 \left[1+\cos\left(\frac{\phi}{F}\right)\right],~~\text{(DE)},\\
V_3(\phi,\chi)&= \frac{\tilde\lambda}{2}\chi^2\phi^2, ~~\text{(DM-DE~interaction)}.           
\end{align} 
One straightforward approach is to solve the Klein-Gordon equations, and then compute 
the densities and other relevant cosmological quantities. However, this approach can be
time consuming in certain regions of the parameter space. Thus a relevant parameter in
the time evolution of the fields is 
the Hubble time $\mathcal{H}^{-1}$ relative to  $m_\chi^{-1}$. For the case when 
 $m_\chi^{-1}\ll \mathcal{H}^{-1}$, one encounters rapid oscillations in the DM field,
 making the computations intractable. To overcome this, it is found 
 preferable to directly solve the differential equations for the density perturbations $\delta_i$
 and the velocity variance $\Theta_i$ averaged over a period of rapid oscillations. For the $\chi$ field, the evolution of the density perturbations is given by
\begin{align}
\delta_\chi^\prime&=\left[3\mathcal{H}(w_\chi-c_{s\chi}^2)-\frac{Q_\chi}{\rho_\chi}\right]\delta_\chi+\frac{3\mathcal{H}Q_\chi}{\rho_\chi(1+w_\chi)}(c_{s\chi}^2-c^2_{\chi_{\rm ad}})\frac{\Theta_\chi}{k}-9\mathcal{H}^2(c_{s\chi}^2-c^2_{\chi_{\rm ad}})\frac{\Theta_\chi}{k}-\Theta_\chi k \nonumber \\
&+\frac{a^2}{k}\frac{\rho_\phi}{\rho_\chi}\Bar{V}_{3,\phi\phi}\Theta_\phi+\frac{1}{\rho_\chi}\Bar{V}_{3,\chi\phi}\phi_0^\prime\chi_1+\frac{1}{\rho_\chi}\bar{V}_{3,\phi}\phi_1^\prime-(3H_L^\prime+kB)(1+w_\chi),
\end{align}
while for the velocity divergence one gets 
\begin{align}
\Theta^\prime_\chi&=(3c_{s\chi}^2-1)\mathcal{H}\Theta_\chi+k\delta_\chi c_{s\chi}^2+3\mathcal{H}(w_\chi-c^2_{\chi_{\rm ad}})\Theta_\chi \non
&~~~-\frac{Q_\chi}{\rho_\chi}\left(1+\frac{c_{s\chi}^2-c^2_{\chi_{\rm ad}}}{1+w_\chi}\right)\Theta_\chi+\frac{k}{\rho_\chi}\Bar{V}_{3,\phi}\phi_1+k(1+w_\chi)A.
\end{align}
In the analysis above two types of sound speeds enter in the equations, i.e.,  $c^2_{s\chi}$ and  $c^2_{\chi_{\rm ad}}$ where $c^2_{s\chi}$ is defined so that
    $c_{s\chi}^2=\delta p_\chi/\delta \rho_\chi$, and  $c^2_{\chi_{\rm ad}}$ 
    depends on the equation of state $w_{\chi}$. Thus    $c^2_{\chi_{\rm ad}}$ 
    is given by
\begin{align}
c^2_{\chi_{\rm ad}}&\equiv\frac{p^\prime_\chi}{\rho^\prime_\chi}=w_\chi-\frac{w^\prime_\chi \rho_\chi}{3\mathcal{H}(1+w_\chi)\rho_\chi-Q_\chi},~~\text{with}~~
     & w_{\chi}^{-1} = 3+ \frac{8 m_\chi^4}{9 \lambda \langle\rho_\chi\rangle}\left(1 + \frac{\tilde{\lambda}\phi_0^2}{m_\chi^2}\right).     
    \label{ichi}
\end{align}
We note that in the absence of self-interactions, $\lambda=0$, which gives
$w_\chi=0$ and CDM is pressure-less as expected. However, in the limit 
when $\dfrac{8m_\chi^4}{9\lambda}\langle\rho_\chi\rangle \ll 1$ one finds $w_\chi\to \frac{1}{3}$ and in this case the $\chi$ field behaves as radiation.
 This is checked numerically in the analysis.

\section{Numerical analysis and fits to cosmological data}

We discuss the numerical analysis of the cosmological parameters 
in two parts. First, we consider a  few benchmarks to study in detail
the effect of  dark energy interaction with dark matter on the cosmological parameters.
In the second part we will run a Markov Chain Monte Carlo analysis and use Bayesian inference to  extract the cosmological parameters. We use the input parameters: $\mu, F, m_\chi, \lambda, \tilde\lambda; \phi_{\rm ini}, \phi'_{\rm ini}; \chi_{\rm ini}, \chi'_{\rm ini}; a_{\rm ini}\sim 10^{-14}$.
Here the background  fields for dark matter and dark energy are 
evolved from early times to late times including the contributions from
neutrinos, baryons, and radiation. The analysis utilizes the 
 Boltzmann solver~\code{CLASS}~\cite{Blas:2011rf} to evolve the background and perturbations equations.  
We have carried out a full numerical analysis where we investigate the effects of varying $\lambda, \tilde \lambda, m_\chi$ on a number of cosmological quantities of interest which consist of $Q_\phi, Q_\chi, \delta_\chi$, $\Theta_\chi,  H(z)$, $w_\phi, w_\chi, \Omega_\phi, \Omega_\chi, \Omega_\gamma, \Omega_b,  P(k),  \frac{\ell(\ell+1)}{2 \pi} C_\ell^{TT}$. 
However, in this note we exhibit only a subset of the results which we next discuss,
and a complete analysis can be found in ref.~\cite{Aboubrahim:2024spa}.

%%%%%%%%%%%%%%%%%%%%%%%%%%%%%%%
\begin{figure}[H]
\begin{centering}
\includegraphics[width=1.0\textwidth]{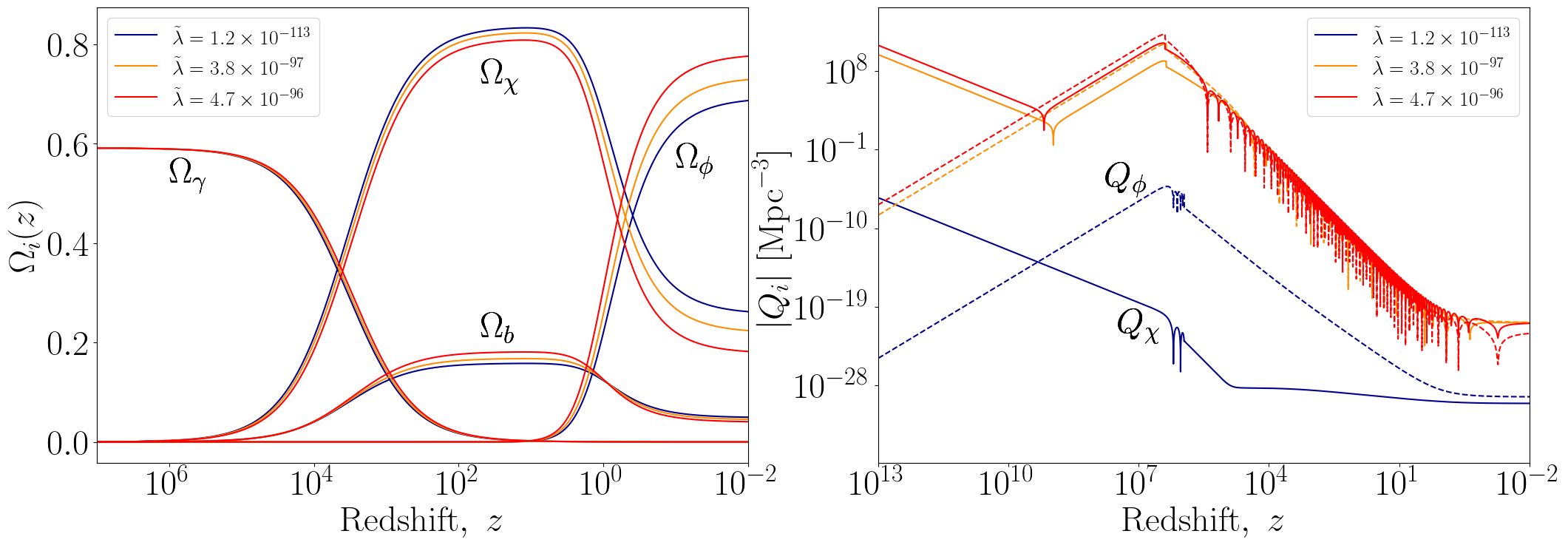}
\caption{  The left panel exhibits the energy density fraction of dark matter, dark energy, baryons and radiation as a function of the redshift for three DM-DE couplings $\tilde \lambda$. The right panel exhibits plots of sources $Q_\phi$ and $Q_\chi$ as a function of the redshift for the same three couplings  of the dark matter-dark energy interaction strengths as the left panel. }
\label{fig1}
\end{centering}
\end{figure}

\begin{figure}[H]
\begin{centering}
\includegraphics[width=1.0\textwidth]{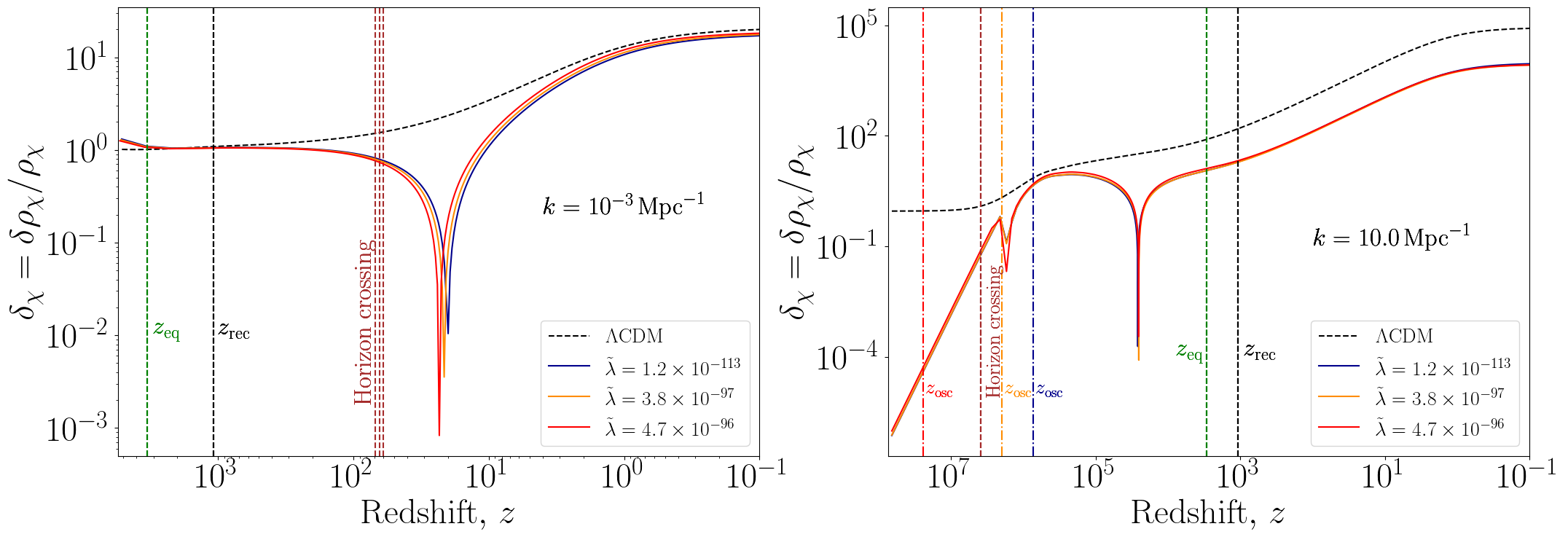} 
\caption{Plots showing the dark matter density perturbations for two values of $k$:  $k=10^{-3}$ Mpc$^{-1}$ (left panel) and $k=10.0$ Mpc$^{-1}$ (right panel), as a function of the redshift $z$ and for three values of $\tilde\lambda$. The three dotted vertical lines correspond to the time of horizon crossing (brown), matter-radiation equality (green) and re-combination (black). The three dash-dot vertical lines correspond to $z_{\rm osc}$, the scale factor when oscillations of the field start, with colors corresponding to each of the three couplings for the dark matter-dark energy interactions.}
\label{fig2}
\end{centering}
\end{figure}

Fig.~\ref{fig1} exhibits the sensitivity of the cosmological parameters on the
 dark matter-dark energy interaction strength $\tilde \lambda$. Here 
 the left panel shows the evolution of the energy density fractions 
$\Omega_\phi, \Omega_\chi, \Omega_\gamma, \Omega_b$ as a function of the
 redshift $z$ from early times to current times. The figure
shows that the energy density fractions $\Omega_\phi$ and $\Omega_\chi$ 
are sensitively dependent on $\tilde\lambda$.  The right panel exhibits
the evolution of the sources $Q_\phi$ and $Q_\chi$  over a wide range of 
redshifts and one finds that $Q_\phi+ Q_\chi\neq 0$ over this entire range
which contradicts the assumption usually made, i.e., that $Q_\phi=- Q_\chi= Q$
in the two-fluid models.
Next we analyze the effect of DM-DE interaction on the density perturbations 
of dark matter $\delta_\chi$ as a function of the redshift $z$. This is exhibited in Fig.~\ref{fig2} for two values of $k$: $k=10^{-3}$ Mpc$^{-1}$ (left panel) and $k=10.0$ Mpc$^{-1}$ (right panel) for the same set of $\tilde \lambda$
as in Fig.~\ref{fig1}. In the left panel one can see that DM-DE
interaction has little effect on the mode at superhorizon scale 
but the effect becomes more prominent after horizon entry where the perturbations corresponding to
different interaction strengths become distinct before increasing and tracing $\Lambda$CDM again. In the right panel, the mode starts to
oscillate as it enters the horizon causing a suppression of growth. Once the mode becomes sub-Jeans, the pressure in the fluid drops and the perturbations grow, trending in the direction of $\Lambda$CDM while remaining suppressed in comparison to CDM.
The dark vertical line marked $z_{\rm rec}$ indicates the point in $z$ where recombination occurs, the red vertical line marked $z_{\rm eq}$ indicated the point of matter-radiation equality, and the blue vertical line indicates the point where the $k$ mode enters the horizon and begins to affect structure formation. 

\begin{figure}[H]
\begin{centering}
\includegraphics[width=1.0\textwidth]{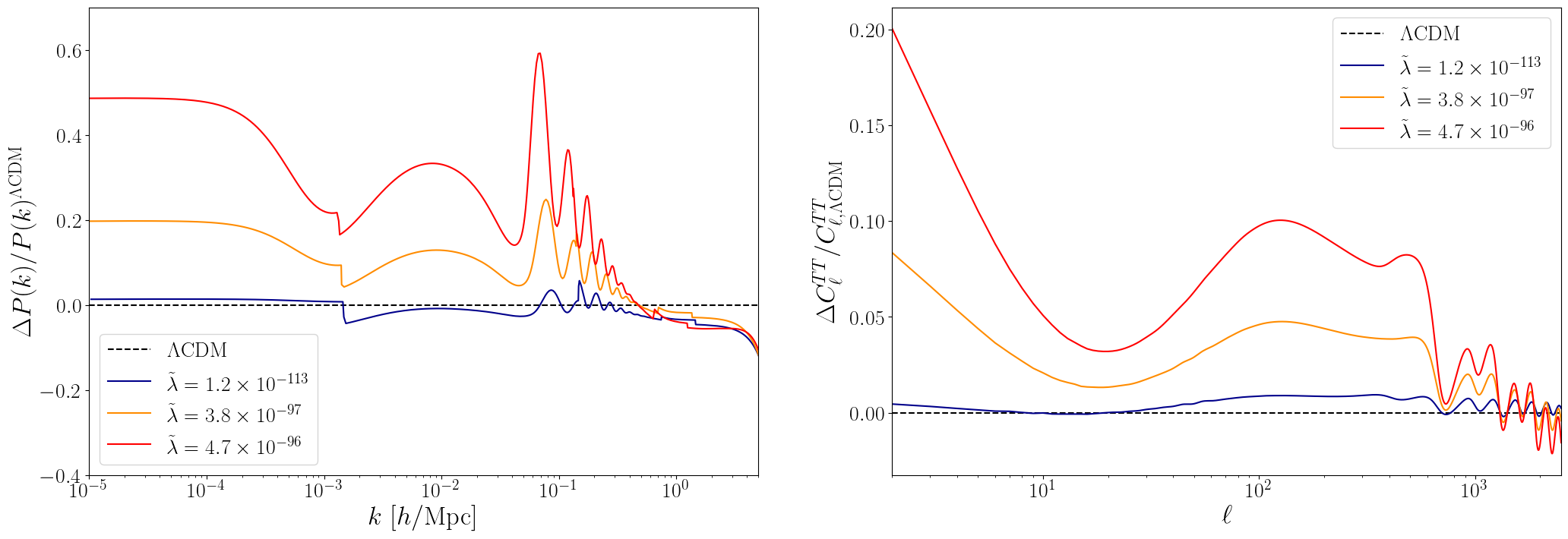}
 \caption{
 Left panel: the relative difference in the matter power spectrum between our model and $\Lambda$CDM plotted against the wavenumber $k$ for three couplings of dark matter-dark energy interactions. Right panel: the relative difference in the temperature power spectrum between our model and $\Lambda$CDM as a function of the multipoles also for three benchmarks of  dark matter-dark energy interactions. The dashed line represents the $\Lambda$CDM model.}
\label{fig3}
\end{centering}
\end{figure}

In the left panel of Fig.~\ref{fig3}, we show the relative difference in the matter power spectrum between our model and $\Lambda$CDM for the three benchmarks.  Here one finds that the effect of
 DM-DE interaction on the matter power spectrum is not significant except 
 for $k<k_{\rm eq}$ which corresponds to large scales.  The right panel of Fig.~\ref{fig3} gives the relative difference in the temperature power spectrum between our model and $\Lambda$CDM as a function of the multipoles also for three benchmarks of DM-DE interactions. Here also one finds that the effect of DM-DE interactions are typically small; specifically the acoustic peak is not much affected relative to the $\Lambda$CDM prediction. However, a significant effect is visible for small $\ell$ values. 
 
%%%%%%%%%%%%%%%%%%%%%%%%%%%%%%%

%%%%%%%%%%%%%%%%%%%%%%%%%%%%%%%
\section{MCMC analysis for  interacting dark matter-dark energy 
using \Q  model}
%%%%%%%%%%%%%%%%%%%%%%%%%%%%%%%
  Before giving our analysis of the cosmological parameters, we discuss briefly some of the tensions appearing in cosmological data.  Thus, currently there is a discrepancy in some observables arising from a mismatch between their inferred values from analysis of the CMB based on $\Lambda$CDM~\cite{Planck:2018nkj,ACT:2020gnv,SPT-3G:2021wgf} and their local direct measurements~\cite{Riess:2019cxk,Wong:2019kwg}.
 Specifically, a $5\sigma$ discrepancy is seen for the Hubble parameter  
 between the Planck collaboration~\cite{Planck:2018vyg} result and the SH0ES collaboration~\cite{Riess:2021jrx} using Cepheid-calibrated supernovae.
 Another discrepancy relates to the clustering of matter at large scales observed
 from galaxy clustering and 
 weak gravitational lensing  surveys~\cite{KiDS:2020suj,Joudaki:2019pmv,DES:2021wwk,Heymans:2020gsg,DES:2020ahh,Kazantzidis:2018rnb} which are seen to be discrepant with the analyses using matter clustering power from the CMB anisotropies based on $\Lambda$CDM. A relevant parameter is $S_8$
 defined as $S_8\equiv\sigma_8\sqrt{\Omega_{\rm m}/0.3}$ which is the weighted amplitude of the variance in matter fluctuations for spheres of size $8h^{-1}$Mpc. Currently $S_8$ shows a
  $2-3\sigma$ tension between the local measurements and the those from CMB using 
    $\Lambda$CDM. Additionally, the most recent results from DESI~\cite{DESI:2024mwx} points to a tension between the DE equation of state (EoS) as predicted by $\Lambda$CDM and that inferred from experiment where the EoS appears to be dynamical. We address the possibility of a dynamical EoS in ref.~\cite{Aboubrahim:2024cyk}.

    %%%%%%%%%%%%%%%%%%%%%%%%
Before going further, we note that prior to the work of ref.~\cite{Aboubrahim:2024spa} many models of DM-DE interaction have been presented.  These include refs.~\cite{Pourtsidou:2013nha,Pourtsidou:2016ico,Linton:2017ged,Chamings:2019kcl,Pan:2019gop,Bonici:2018qli,Yang:2019uzo,Pan:2020zza,Wang:2016lxa,Wang:2024vmw,Bamba:2012cp}.
Some of the works have used variational approach to introduce the couplings~\cite{Boehmer:2015kta,Boehmer:2015sha,Archidiacono:2022iuu,Rezazadeh:2022lsf} 
while other works have used interactions at the level of continuity equations for the energy densities treating both DM and DE as fluids~\cite{Potter:2011nv,DiValentino:2019jae,DiValentino:2019ffd,Escamilla:2023shf,Bernui:2023byc,Sharma:2021ayk,Sharma:2021ivo,Yang:2022csz}, DM is a fluid while DE is quintessence~\cite{Amendola:1999er,Kase:2019veo,Perez:2021cvg,Lee:2022cyh} and both DM and DE are scalar fields~\cite{Garcia-Arroyo:2024tqq,Beyer:2010mt,Beyer:2014uqa,vandeBruck:2022xbk}.
 A recent review of several models can be found in~\cite{Abdalla:2022yfr}. 
 
\Q  is different from previous works in that it is fully field-theoretic with no 
    extraneous ad hoc assumptions made in fluid equations. We give now the result of our analysis within \Q. The data sets used in our analysis are as follows:
the Planck data on anisotropies and polarization measurements~\cite{Planck:2018vyg,Planck:2018nkj,Planck:2019nip} and Planck lensing data~\cite{Planck:2018lbu}.
For Baryon Acoustic Oscillation (BAO) the data sets from the Sloan Digital Sky Survey is used which includes several surveys~\cite{Beutler:2011hx,BOSS:2012bus,Howlett:2014opa,Ross:2014qpa,BOSS:2016wmc,eBOSS:2020yzd}.
 For Pantheon+SH0ES the data sets are from~\cite{Brout:2022vxf,Riess:2021jrx}. 
 For  WiggleZ, the Large Scale Structure data survey is used~\cite{Parkinson:2012vd}. 
Using the data sets above we perform a MCMC fit to the cosmological data in five different combinations and we look for the best fits to the cosmological parameters of our model while examining the effect they have on other important parameters such as: $H_0, \Omega_m, \Omega_\phi, \sigma_8, S_8$.
To check the goodness of the fits we define:
\begin{align}
 \Delta \chi^2 _{\text{min}} =
 \chi^2 _{\text{min}, {\mathcal Q}_{\rm CDM}}- \chi^2 _{\text{min}, \Lambda\text{CDM}}.
 \end{align}
The result of the analysis is exhibited in Tables~\ref{tab1}$-$\ref{tab3}.

%%%%%%%%%%%%%%%%
\begin{table}[H]
\centering
{\tabulinesep=1.2mm
\begin{tabu}{cc}
\hline\hline
Data sets/Theory & $\Delta \chi^2 _{\text{min}}$    \\
\hline\hline
\text{Planck + BAO} & 0.0\\ 
  \text{Planck+ Lensing}  & 0.0\\   
 \text{Planck + Pantheon + SH0ES } & $-1.0$ \\ 
   \text{Planck+ Lensing + BAO+ WiggleZ} &  $+1.0$ \\   
  \text{All data sets} &  $-1.0$ \\
\hline\hline
\end{tabu}}
\caption{Comparison of \Q analysis with that of $\Lambda$CDM.}
\label{tab1}
\end{table}

\begin{table}[H]
\centering
{\tabulinesep=1.2mm
\begin{tabu}{cc}
\hline\hline
Data sets/Theory & $H_0$    \\
\hline\hline
 \text{SH0ES}  &$H_0^{\rm R22}=(73.04\pm 1.04) \text{km/s/Mpc}$\\
     \text{Planck}  &$H_0^{\rm Pl}=(67.4\pm 0.5) \text{km/s/Mpc}$\\
\Q  &  $H_0=(68.84_{-0.24}^{+2.10})\text{ km/s/Mpc}$\\ 
\hline\hline
\end{tabu}}
\caption{Comparison of \Q analysis for $H_0$ with those of  Planck and SH0ES.}
\label{tab2}
\end{table}

In Table~\ref{tab1}, we give a comparison
of the goodness of \Q fits relative to that of $\Lambda$CDM. Here
the first two data sets show no difference between \Q and 
$\Lambda$CDM. The third data set and the combination of all data show that  \Q
 fits the data better, although only slightly as exhibited in Table~\ref{tab1}.
 In Table~\ref{tab2}, we give a  comparison between the values of $H_0$  in our \Q model and those obtained by 
   Planck and SH0ES.  Here the $H_0$ tension in $\Lambda$CDM with R22 is more than 5$\sigma$, while the $H_0$ from our analysis based on \Q is now $\sim 2.7 \sigma$ away from the R22 measurement indicating a slight improvement in reducing the tension.
In Table~\ref{tab3},  we discuss the $S_8$ tension.  Here one finds that the
   \Q analysis (using the Planck + Pantheon + SH0ES data sets) is consistent with both KiDS and DES, and resolves the $\sim 3\sigma$ tension that $S_8$ has with the Standard Model.
A similar result in resolving the $S_8$ tension is based on including a drag term between DM and DE is discussed in ref.~\cite{Poulin:2022sgp}.

%%%%%%%%%%%%%%%%%

\begin{table}[H]
\centering
{\tabulinesep=1.2mm
\begin{tabu}{cc}
Data sets/Theory & $S_8$\\
\hline\hline
 \text{Planck}  &$S_8^{\rm Pl}=0.834\pm 0.016$\\
   \text{KiDS-1000} &   $S_8^{\rm KiDS}=0.759_{-0.021}^{+0.024}$\\
   \text{DES-Y3}  &$S_8^{\rm DES}=0.759_{-0.023}^{+0.025}$\\
    \Q & $S_8=0.7975_{-0.0250}^{+0.0180}$ \\
\hline\hline
\end{tabu}}
\caption{Comparison of \Q analysis for $S_8$ with those of  Planck,
KiDS-1000 and DES-Y3.}
\label{tab3}
\end{table}

%%%%%%%%%%%%%%%%%%

\section{Conclusion}

The two-fluid model for dark matter and dark energy which uses an ad hoc assumption of an interaction
between them at the level of the continuity equations, does not arise from an underlying Lagrangian and is not at the same footing as the Standard Model of particle physics or Einstein's gravity. We have discussed an alternative approach, i.e., \Q,  which is field-theoretic and produces a consistent set of continuity equations for dark matter and dark energy replacing the fluid equations currently is use.
 Thus the \Q model
  provides the proper framework for  cosmological analyses. We have carried out fits to the cosmological data using \Q
  and find that the $\chi^2$ of our fits to be at the same level as the $\Lambda$CDM. Observables such as the Hubble parameter and $S_8$ are found to be sensitive to dark matter self-interaction as well as to DM-DE interactions and this helps alleviate the tension for $H_0$ while resolving the $S_8$ tension for some data sets. \Q  is theoretically robust and with more data we should be able to
  discriminate further \Q from $\Lambda$CDM. Finally, in addition
to the work discussed above, a new analysis discusses a new phenomenon related to the
transmutation of dark energy from thawing to scaling freezing in the late universe~\cite{Aboubrahim:2024cyk}.

\acknowledgments
The research of PN was supported in part by the NSF Grant PHY-2209903.

\bibliographystyle{JHEP}
\bibliography{references}

\end{document}